\documentclass[twocolumn,amsmath,amssymb,prl,nofootinbib,10pt,superscriptaddress]{revtex4-1}

\usepackage[english]{babel}
\usepackage{amsmath}
\usepackage{graphicx, xcolor}
\usepackage[colorlinks=true, allcolors=blue]{hyperref}
\usepackage{ulem}
\usepackage{float}

\newcommand{\calR}{{\cal R}}
\newcommand{\Up}{{U_{\phi}}}
\newcommand{\Upp}{{U_{\phi\phi}}}

\newcommand{\DeltaN}{N}
\newcommand{\dd}{\text{d}}

\begin{document}
{\small{\textsc{Matches published version in Phys. Rev. L.}}}
\title{Extended $\delta N$ Formalism: Nonspatially Flat Separate Universe Approach}

\author{Danilo Artigas}%
\email{artigas@tap.scphys.kyoto-u.ac.jp}
\affiliation{%
Department of Physics, Kyoto University, Kyoto 606-8502, Japan
}%

\author{Shi Pi}%
\email{shi.pi@itp.ac.cn}
\affiliation{%
Institute of Theoretical Physics,
Chinese Academy of Sciences, Beijing 100190, China
}%
\affiliation{%
Center for High Energy Physics, Peking University, Beijing 100871, China
}%
\affiliation{%
Kavli Institute for the Physics and Mathematics of the Universe (WPI), UTIAS, 
The University of Tokyo, Kashiwa, Chiba 277-8583, Japan
}%

\author{Takahiro Tanaka}%
\email{t.tanaka@tap.scphys.kyoto-u.ac.jp}
\affiliation{%
Department of Physics, Kyoto University, Kyoto 606-8502, Japan
}%
\affiliation{%
Center for Gravitational Physics and Quantum Information, Yukawa
Institute for Theoretical Physics, Kyoto University, Kyoto 606-8502, Japan
}%

\begin{abstract}
The $\delta N$ formalism is a powerful approach to compute nonlinearly the large-scale evolution of the comoving curvature perturbation $\zeta$. It assumes a set of FLRW patches that evolve independently, but in doing so, all the gradient terms are discarded, which are not negligibly small in models beyond slow roll.
In this Letter, we extend the formalism to capture these gradient corrections by encoding them in a homogeneous-spatial-curvature contribution assigned to each FLRW patch. For a concrete example, we apply this formalism to the ultra-slow-roll inflation, and find that it can correctly describe the large-scale evolution of the comoving curvature perturbation from the horizon exit. We also briefly discuss non-Gaussianities in this context.
\end{abstract}
\maketitle

\textit{Introduction}---The curvature perturbation on comoving slices, $\zeta$, is the seed of cosmic microwave background anisotropies and large-scale structures, which originate from the quantum fluctuations of the inflaton field stretched out of the Hubble horizon during inflation. On superhorizon scales, the evolution of the curvature perturbation can be well described by the $\delta N$ formalism \cite{Lifshitz:1960, Starobinsky:1982ee, Salopek:1990jq, Comer:1994np, Sasaki:1995aw, Sasaki:1998ug, Wands:2000dp, Lyth:2003im, Rigopoulos:2003ak, Lyth:2004gb,Lyth:2005fi}, which is based on the fact that the distant Hubble patches evolve independently, i.e., via the separate-universe approach. In this picture, quantum fluctuations exiting the Hubble horizon are described as a classical field, homogeneous on each patch but with different values on each causally disconnected Hubble patch. These patches evolve independently on superhorizon scales until the end of inflation and the local expansion of each patch is described by the $e$-folding number $N$. The usual $\delta N$ formalism tells us that the curvature perturbation $\zeta$ on the final comoving hypersurface of a Hubble patch is given by the difference in the $e$-folding number between its local expansion and the fiducial one, i.e., $\zeta=\delta N$. 
This simple formula is very useful in various inflation models,
such as ultra-slow-roll inflation \cite{Namjoo:2012aa,Chen:2013eea,Cai:2018dkf,Pattison:2018bct,Pi:2022ysn}, constant-roll inflation \cite{Atal:2018neu,Atal:2019cdz,Escriva:2023uko,Wang:2024xdl} or the curvaton scenario \cite{Sasaki:2006kq,Fujita:2014iaa,Ando:2017veq,Pi:2021dft,Chen:2023lou}. 
Also, it can be applied to the stochastic approach \cite{Fujita:2013cna, Fujita:2014tja, Vennin:2015hra, Pattison:2019hef, Pattison:2021oen, Briaud:2023eae}.

Recently, it was shown that the separate-universe approach transiently breaks down around the slow-roll-to-ultra-slow-roll transition \cite{Leach:2001zf, Naruko:2012fe, Domenech:2023dxx, Jackson:2023obv}. This is mainly because of the nonnegligible superhorizon evolution of $\zeta$, which at the leading order is dominated by the spatial-gradient term and gives the behavior of the power spectrum $\mathcal{P}_\zeta\propto k^4$~\cite{Leach:2001zf,Byrnes:2018txb,Cole:2022xqc}. One way of solving this problem is to wait and apply the $\delta N$ formalism only at a later time $t_j$ when the spatial-gradient terms are again negligible. The price we have to pay is to solve the linear perturbation equation without neglecting spatial gradient up to this moment $t_j$, which can be more than a few $e$-folds later than the horizon-exit time $t_k$, 
and the convenience of $\delta N$ formalism is significantly lost.

In this Letter, we propose a novel improvement to the separate-universe approach and an extended $\delta N$ formalism by taking into account the local spatial scalar curvature.
This is a different direction from the anisotropic extensions given in Refs.~\cite{Abolhasani:2013zya, Talebian-Ashkezari:2016llx, Talebian-Ashkezari:2018cax, Tanaka:2021dww,Tanaka:2023gul}.
In this new framework, the separate universe approximates each Hubble patch as a local homogeneous and isotropic Friedmann-Lemaître-Robertson-Walker (FLRW) universe with a curvature term, which can be evolved without referring to the adjacent patches. We will show that this extended $\delta N$ formalism which takes the spatial curvature of each FLRW patch into account can correctly describe the superhorizon evolution of $\zeta$:
even setting the initial time at the horizon-exit moment $t_k$, we obtain an accurate power spectrum that fits the numerical results quite well. This implies that our extended $\delta N$ formalism can be safely applied to cases where the evolution significantly deviates from the slow-roll attractor, such as ultra-slow-roll inflation. 


\textit{Extended $\delta N$ formalism}---We work with the perturbed spatial metric of the scalar-type \cite{Bardeen:1980kt,Kodama:1984ziu,Sasaki:1998ug} 
\begin{align}\nonumber
ds^2_{(3)}=&\, a^2 e^{2\mathcal{R}}
\left(\delta_{ij}-2{\nabla^{-2}}\nabla_i \nabla_j H_T\right) \dd x^i\dd x^j,\nonumber
\end{align}
where $a$ is the scale factor, $\cal R$ is the nonlinear curvature perturbation and $H_T$ denotes the traceless part of metric perturbations.
The gauge-invariant curvature perturbation on comoving slices $\zeta$ is defined linearly by \cite{Mukhanov:1988jd,Sasaki:1986hm}
\begin{equation}
 \zeta:=\calR- \frac{a H}{\phi'} {\delta\phi}\,, 
\label{defchiF}
\end{equation}
where $H$ is the Hubble expansion rate, and we denote by a prime the differentiation in the conformal time, $\eta\equiv\int\dd t/a(t)$. $\delta\phi$ is the perturbation of the inflaton field $\phi$ in this arbitrary gauge. At linear order, $\zeta$ satisfies the following Mukhanov-Sasaki equation
\begin{equation}
 \zeta''+2\frac{z'}{z}\zeta'+k^2\zeta=0\,, \label{zetaeq}
\end{equation}
with $z\equiv\phi'/H$ and $k$ the wave number. 
Equation~\eqref{zetaeq} has a trivial solution $\zeta=\text{constant}$ at the leading order of $k^2$, which is known as the adiabatic mode. 
By solving equation \eqref{zetaeq} numerically, we can get the exact result of $\zeta$ in linear-perturbation theory.

Late-time evolution can also be achieved by the $\delta N$ formalism, which is based on the superhorizon solution of \eqref{zetaeq} with $k^2\to0$. However, in some models, the $k^2$ term may not be negligible right at the horizon exit \cite{Domenech:2023dxx, Jackson:2023obv}. Here we will first show that the $\mathcal{O}(k^2)$ correction is important in ultra-slow-roll inflation, and then propose an extended $\delta N$ formalism to take the $k^2$ term into account. Equation~\eqref{zetaeq} has the formal solution
\begin{align}
 \zeta= \zeta\left(\eta_{\rm ref}\right) u_{\rm ad}(\eta)+\zeta'\left(\eta_{\rm ref}\right) u_{\rm nad}(\eta)\,, \nonumber
\end{align}
where the adiabatic and nonadiabatic mode functions are
\begin{align}\nonumber
    u_\text{ad}(\eta)&= 1-k^2\int_{\eta_{\rm ref}}^{\eta}\frac{d\eta'}{z^2(\eta')}
                       \int_{\eta_{\rm ref}}^{\eta'}{d\eta''}{z^2(\eta'')}+\cdots,\\\nonumber
    u_\text{nad}(\eta)&= z^2(\eta_{\rm ref})\int_{\eta_{\rm ref}}^{\eta}\frac{d\eta'}{z^2(\eta')}\\\nonumber
    &\quad\times\left(1-k^2\int_{\eta_{\rm ref}}^{\eta'}{d\eta''}{z^2(\eta'')}
                     \int_{\eta_{\rm ref}}^{\eta''}\frac{d\eta'''}{z^2(\eta''')} +\cdots\right). 
\end{align}
Here, $\eta_{\rm ref}$ is an arbitrary reference time. The solutions $u_{\text{ad}}$ and $u_{\text{nad}}$ are degenerate in the sense that a part of the leading order of $u_\text{nad}$ can be transferred to the subleading order term in $u_\text{ad}$ by changing the initial time $\eta_{\rm ref}$.  Furthermore, when $z$ is rapidly decreasing, 
the next-to-leading $k^2$ correction of $u_\text{nad}$ is in general suppressed on superhorizon scales, which does not give any growth in the later stage of inflation. On the other hand, the gradient term of the adiabatic counterpart is not always suppressed, which requires an accurate treatment even on superhorizon scales. This is our main motivation to propose the extended $\delta N$ formalism.

\begin{figure}[t]
\includegraphics[width=0.48\textwidth]{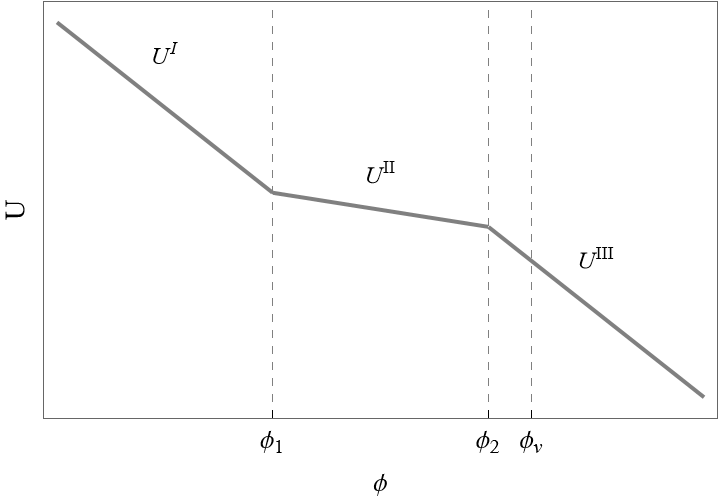}
\caption{Schematic view of the Starobinsky potential.
}
\label{fig:potential}
\end{figure}

As a simple example, we consider the Starobinsky's linear potential model~\cite{Starobinsky:1992ts,Ivanov:1994pa,Biagetti:2018pjj,Ozsoy:2019lyy,Pi:2022zxs}, in which the potential $U\left(\phi\right)$ is piecewise linear, i.e., the potential slope $U_\phi$ is constant in each region given by 
\begin{equation}\label{def:potential}
 \Up=\left\{
 \begin{array}{ll}
  U^{\rm I}_\phi  \,,\qquad &(\phi<\phi_1,~~\mbox{segment I})\,,\cr
  U^{\rm II}_\phi \,,\qquad & (\phi_1<\phi<\phi_2,~~\mbox{segment II})\,,\cr
    U^{\rm III}_\phi  \,,\qquad & (\phi_2<\phi,~~\mbox{segment III})\,,\cr
  \end{array}\right.
\end{equation}
with  $|U^{\rm I}_\phi|=|U^{\rm III}_\phi|\gg |U^{\rm II}_\phi|$. The potential is represented in Fig.~\ref{fig:potential}. From now on, we use the $e$-folding number $N$ as the time coordinate. We denote the transition time of the potential slope by 
$N_i$, i.e., $\phi_i\equiv \phi\left(N_i\right)$ for $i=1, 2$.
The system undergoes a slow-roll evolution until $\phi=\phi_1$. After the transition at $\phi_1$, the velocity inherited from segment I is much larger than the attractor velocity on the shallower potential of segment II, which leads to a violation of slow-roll condition for a few $e$-folds. This is called the ultra-slow-roll phase, during which the inflaton continues to decelerate and finally reaches the attractor of segment II at $\phi_v$. To avoid lasting inflation in such a flat potential $U^{\rm II}$ for too long, we cut and match it to another steeper slow-roll potential $U^{\rm III}$ at $\phi_2<\phi_v$. The sudden change of the slope at $\phi_2$ guarantees that the contribution of $\delta N$ to $\zeta$ is mainly due to the ultra-slow-roll stage, thus the non-Gaussianity is large \cite{Pi:2022ysn}.

We assume that the evolution of a $k$-mode can be described by the linear-perturbation theory until some time $N_j$. After horizon-crossing time $N_k$, the initial conditions for the separate-universe evolution at $N_j(\geq N_k)$ are set by the linear-perturbation theory, and the 
superhorizon evolution after $N_j$ can be described by the $\delta N$ formalism. Of course, if we set $N_j$ to be the end of inflation $N_\mathrm{end}$, the numerical solution of the Mukhanov-Sasaki equation~\eqref{zetaeq} will give the accurate linear curvature perturbation. In the standard $\delta N$ approach, for the separate universe to be accurate, one should wait until $N_{kj}\equiv N_j-N_k$ is large enough. For slow-roll inflation, a few $e$-folds can work perfectly. However, as we mentioned above, in the ultra-slow-roll inflation, for some wave numbers which exit the horizon around the slow-roll-to-ultra-slow-roll transition ($N_k\sim N_1$), we need to set $N_{kj}$ more than a few~\cite{Leach:2001zf,Jackson:2023obv,Domenech:2023dxx}, and the $\delta N$ formalism loses its convenience. However, in the extended $\delta N$ formalism that we propose below, the result is quite accurate even for $N_{kj}\approx 0$, because it takes into account the spatial curvature of the foliation in its initial condition.

For simplicity, we adopt the de Sitter approximation, which fixes the energy density to a constant, $3H_0^2$, i.e., the energy density is dominated by the constant part of the inflaton potential. For an arbitrary patch with curvature, the expansion rate is given by
\begin{equation}\label{localFriedmann}
    H^2=H_0^2-{\cal K}e^{-2(N-N_j)}\,, 
\end{equation}
where ${\cal K}$ represents the spatial 
curvature evaluated at the junction time $N_j$, and in this Letter, we will neglect all terms of order $\mathcal{O}\left(\mathcal{K}^2\right)$ and beyond. 
Note that, unlike the usual Friedman equation for the entire universe, $\mathcal{K}e^{2N_j}$ cannot be normalized to $\pm1$, as we do not have degrees of freedom to adjust the scale factor $a=e^N$ according to $\mathcal{K}$ that varies in different patches. Keeping in mind that $dN=Hdt$, 
the homogeneous scalar field in a spatially-curved patch obeys the following Klein-Gordon equation, which can be derived from Eq.~\eqref{bgNK} in the \textit{end matter} up to $\mathcal{O}(\mathcal{K})$:
\begin{align}\label{KG(phi)}
  &\left[\partial_N^2+ 3\partial_N\right]\phi +\frac{\Up}{H_0^2} + \frac{{\cal K}}{H_0^2} e^{-2(N-N_j)}\left(\phi_N+ \frac{\Up}{H_0^2}\right)=0\,, 
\end{align}
where $\phi_N\equiv \partial_N\phi$. 
We expand $\phi$ in powers of ${\cal K}$ as $\phi=\phi^{(0)}+\phi^{(1)}+\cdots$.
At the lowest order in ${\cal K}$, we have the usual second-order differential equation, of which the solution in each segment is
\begin{align} \nonumber
\phi^{(0)}(N)&=\phi^{(0)}(N_*) -\frac{\Up}{3H_0^2} (N-N_*) \\\label{phi0}
&-\frac13\left(\phi^{(0)}_N(N_*)+\frac{\Up}{3 H_0^2}\right) \left( e^{-3(N-N_*)}-1\right) \,,\end{align}
for the initial conditions set at $N=N_*$, which refers either to $N_j$ or to $N_1$ depending on which segment is concerned and the value of $N_k$. 
It is straightforward to obtain the equation of motion for $\phi^{(1)}$ which contains the ${\cal{O}}({\cal K})$ correction, 
\begin{align}
    \left[\partial_N^2+3\partial_N\right]\phi^{(1)}
   &= - \frac{{\cal K}}{H_0^2} e^{-2(N-N_j)}\left[
      \phi_N^{(0)} +\frac{\Up}{H_0^2}\right] \,,\nonumber
\end{align}  
and the solution at this order is
\begin{align} \label{phi1}
\phi^{(1)}(N)&=\phi^{(1)}(N_*) - \frac{1}{3} \phi^{(1)}_N(N_*)\left(e^{-3\left(N-N_*\right)}-1\right) \cr
    &+ \frac{\mathcal{K} U_\phi}{3 H_0^4} e^{-2\left(N-N_j\right)} \left[ - \frac{2}{5} e^{-2 \left(N_* - N\right)} + 1 \right. \cr
    &\left.- \frac{1}{2} e^{-\left(N-N_*\right)} - \frac{1}{10} e^{-3\left(N-N_*\right)} \right] \nonumber \\
    & - \frac{\mathcal{K} \phi^{(0)}_N (N_*)}{6 H_0^2} e^{-2\left(N-N_j\right)} \left[ \frac{2}{5} e^{-2 \left(N_* - N\right)} \right. \cr
    &\left.-  e^{-\left(N-N_*\right)} + \frac{3}{5} e^{-3\left(N-N_*\right)} \right] \,. 
\end{align}
Then $\phi_N^{(0)}(N)$ and $\phi_N^{(1)}(N)$ can be calculated by taking the derivative of \eqref{phi0} and \eqref{phi1}.

Knowing the evolution of the inflaton field $\phi(N)$ up to $\mathcal{O}(\mathcal{K})$, what we want to calculate is the comoving curvature perturbation $\zeta(N)$ at a late time. In the model considered in \eqref{def:potential}, as the transition from ultra-slow roll to the second slow-roll stage is abrupt, the contribution to $\delta N$ from stage III is negligible \cite{Chen:2013aj,Cai:2018dkf,Pi:2022ysn,Pi:2024jwt}. Therefore, the curvature perturbation will not change much after $N_2$ and the constant-$\phi$ hypersurface at $\phi_2$ is chosen as the final comoving slice, i.e., $\zeta(N_\mathrm{end})\approx\zeta(N_2)$, in the analytic calculation below. 
The general methodology to solve the dynamics is the following. \\

\noindent
(a) We choose the $\delta N$ gauge, in which the shift vanishes and $\mathcal{R}'=H_T'/3$~\cite{Sasaki:1998ug}. In this gauge,
the physical volume is proportional to $\exp\left(3N\right)$, independently of the spatial coordinates. At $N=N_j$, we set the initial conditions of the field perturbation $\delta\phi$ and the curvature perturbation $\mathcal{R}$ for the $\delta N$ formalism to match the linear-perturbation theory. The results are 
\begin{align}
    &\delta\phi\left(N_j\right)=0\,,\qquad \mathcal{R}\left(N_j\right)=\zeta\left(N_j\right)\,, \cr
    &\delta\phi_N\left(N_j\right)=\frac{\phi_N}{3H_0^2}\left(-U
    \zeta_N\left(N_j\right)+k^2 e^{-2N_j}\zeta\left(N_j\right)\right) \,, \cr
    &\mathcal{R}_N\left(N_j\right) = \frac{\phi_N^2 }
    {6}
    \zeta_N\left(N_j\right) +\frac{k^2}{3H_0^2}e^{-2N_j}\zeta\left(N_j\right) \,, \cr
    & e^{2N_j} \mathcal{K} = \frac{2k^2}{3} \zeta\left(N_j\right) \,. \label{initConds}
\end{align}
The derivation of these initial conditions is shown in details in the \textit{end matter}. As we mentioned before, in the present model we can safely adopt the de Sitter assumption and replace $H$ by $H_0$. The first line comes from setting the initial surface at $N_j$ in the comoving slicing $\delta\phi=0$. This choice is allowed because of the existence of residual gauge freedom in the $\delta N$ gauge, which corresponds to the choice of the origin of time coordinate in each local universe \cite{Artigas:2021zdk,Artigas:2023kyo}. 

In such a separate universe, the spatial gradient of the scalar field is absent in the Klein-Gordon equation of $\phi$, Eq.~\eqref{KG(phi)}. To ensure that the equation of motion for $\delta\phi$ matches with the Klein Gordon equation \eqref{KG(phi)} in the perturbed universe, we set $\delta \phi=0$ at $N=N_j$ using the residual gauge degree of freedom. The first term in 
$\delta\phi_N(N_j)$ is the contribution from the nonadiabatic mode $u_{\rm nad}$, while the second term is $\mathcal{O}\left(k^2\right)$. Hence the corrections to $\delta\phi$ from gradient terms can be neglected. The second and third lines in Eqs.~\eqref{initConds} are obtained with the aid of the momentum constraint.
In the last line, the effective curvature $\propto\mathcal{K}$ in the separate-universe approach 
is determined by the term-by-term matching of perturbed \eqref{KG(phi)} and the equation of motion for $\delta\phi$
at the linear order. 
Equivalently, we can also get this relation by evaluating the spatial Ricci curvature on the $\delta\phi=0$ hypersurface. While the value of $k^2$ is necessary to give the initial condition for 
the long-wavelength perturbations, the long-wavelength evolution itself is completely local.
In $\delta N$ gauge, the equation for $\mathcal{R}$ valid up to $\mathcal{O}(k^2)$ is linear and closed as 
\begin{equation}
 \frac1{a^3 H_0} \partial_N \left(a^3 H_0 \partial_N{\cal R}\right)
   =\frac{k^2}{3a^2 H_0^2}{\cal R} \,. 
\end{equation}
The solution up to $\mathcal{O}\left(k^2\right)$ under the initial conditions~\eqref{initConds} is obtained by a perturbative expansion in $k^2$ as 
\begin{equation}
\mathcal{R}(N)= \zeta\left(N_j\right) + \frac{k^2 \zeta\left(N_j\right)}{6 H_0^2} \left(e^{-2N_j} - e^{-2N} \right) \,. 
\end{equation}
The second term gives a weak time dependence, which anyway remains minor.

\noindent
(b) To evaluate the $e$-folding number, we only need to solve the background equations of motion \eqref{bgNK}--\eqref{frNK} in the \textit{end matter}. For our specific example of potential shown in Fig.~\ref{fig:potential}, this can be done in the following way:

(b.1)  If $N_j$ is in the segment I, the evolution in the segment I is given by Eqs.~\eqref{phi0}--\eqref{phi1} and their derivatives, setting $N_*=N_j$. We solve the fields up to the transition at $\phi=\phi_1$ which provides the initial conditions for the succeeding ultra-slow-roll evolution.

(b.2) For the field evolution in segment II, one uses again equations~\eqref{phi0}--\eqref{phi1} and their derivatives but with the initial conditions at $N_*\equiv N_1$ for $N_j<N_1$ and those at $N_*\equiv N_j$ for $N_j>N_1$.

(b.3) The numbers of $e$-folds in the respective segments $N_{ab}\equiv N_b- N_a$ are obtained by inverting Eqs.~\eqref{phi0} and~\eqref{phi1}. They depend on $\phi_{ab}\equiv \phi_b-\phi_a$ as
\begin{align}
    N_{j1} =& \frac{3 H_0^2}{U_{\phi}^{\rm I}} \left[- \phi_{j1} + \frac{1}{3}\left(\phi_N(N_j) + \frac{U_{\phi}^{\rm I}}{3 H_0^2}\right)\right. \cr
    &\qquad \left. - \frac{2\mathcal{K} U_\phi^{\rm I}}{15 H_0^4} - \frac{\mathcal{K} \phi_N^{(0)}}{15 H_0^2} \right] 
 \,, \label{Nj1} \\
    N_{12} =& \frac{3 H_0^2}{U_\phi^{\rm II}} \left[ - \phi_{12} - \frac{1}{3} \left(\phi_N(N_1) + \frac{U_\phi^{\rm II}}{3 H_0^2}\right) \left(e^{-3 N_{12}}-1\right) \right.\cr
    &\qquad \left. - \frac{2 \mathcal{K} U_\phi^{\rm II}}{15 H_0^4} e^{-2 N_{j1}} - \frac{\mathcal{K} \phi_N^{(0)}(N_1)}{15 H_0^2} e^{-2 N_{j1}} \right]
\,, \label{N12}
\end{align}
where we have neglected the remaining terms that decay exponentially fast.
For a detailed comparison with the ordinary linear perturbation, see Supplemental Material.

\noindent
(c) We evaluate $\zeta$, the nonlinear curvature perturbation on a final comoving hypersurface $N_2=\bar{N}_2+\delta N_2$ where $\phi=\phi_2$ and we now denote background quantities with an overline. Since the initial surface is in the $\delta N$ gauge, $N_j=\bar{N}_j$ and we can write $\delta N_2 = \delta N_{j2}$. We then perform the gauge transformation $N_2\rightarrow N_2^{(c)}=N_2-\delta N_{j2}$ such that $\phi^{(c)}(N_2^{(c)})=\bar{\phi}^{(c)}(\bar{N}_2)=\phi_2$, where the superscript $(c)$ denotes variables in the comoving slicing. Since our gauge transformation doesn't change spatial coordinates, the local expansion on the final hypersurface $\exp\big[N_2+\mathcal{R}(N_2)\big]$ is conserved. Plugging the gauge transformation, one finds the transformation for $\mathcal{R}$, namely,
\begin{align}\label{maineqn}
    \zeta(\bar{N}_2) &:= \mathcal{R}^{(c)}(\bar{N}_2) = \mathcal{R}\left(N_2\right) + \delta N_{j2}\,. 
\end{align} 
Since at a late epoch $\mathcal R$ is constant we can approximate $\mathcal{R}(N_2)\approx \mathcal{R}(\bar{N}_2)$ \cite{Sasaki:1998ug}.
As the difference between $\mathcal{R}(N_2)$ and $\mathcal{R}(N_j)$ remains small, we can also approximate \eqref{maineqn} by $\zeta(\bar{N}_2)=\mathcal{R}(N_j)+\delta N_{j2}$, which is closer to the ordinary $\delta N$ formula.

\begin{figure}[t]
\includegraphics[width=0.48\textwidth]{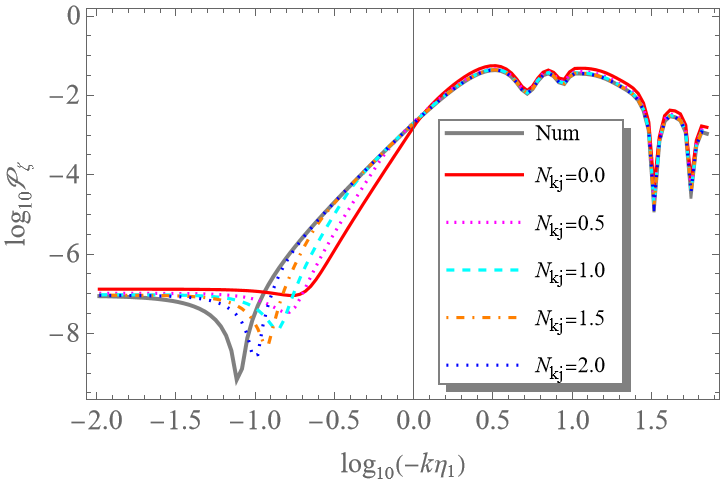}
\vspace{1em}
\includegraphics[width=0.48\textwidth]{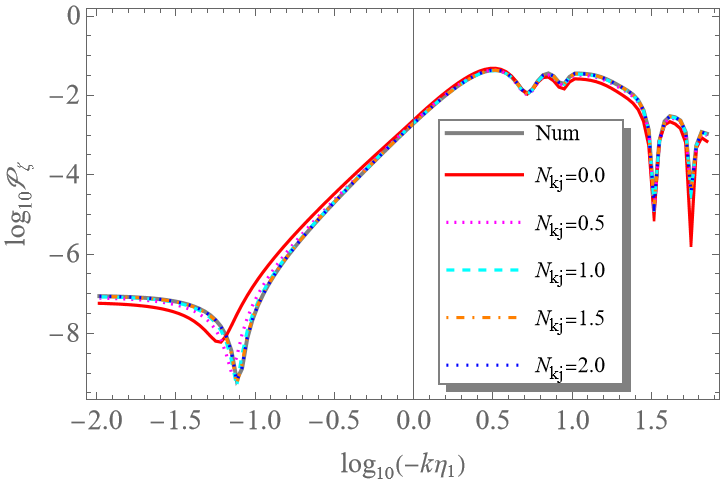}
\caption{Power spectrum generated using the standard $\delta N$ (top) and the extended $\delta N$ formalism (bottom). Here, the parameters were fixed to $H_0=10^{-6}$, $U_\phi^{\rm I}=U_\phi^{\rm III}=-10^{-14}$, $U_\phi^{\rm II}=0$ and $\phi_{12}\approx 0.0011$ such that $N_{12}=2$.
``Num" means the results obtained by solving numerically the Mukhanov-Sasaki equation until the end, and $N_{kj}=0.0$, 0.5, etc., means using different initial times in the two $\delta N$ formalisms.
}
\label{fig:SpecModel2}
\end{figure}

The power spectrum of $\zeta$ can then be calculated under this formalism. When the usual separate universe is matched to perturbation theory right after the horizon exit of the $k$ mode, $N_{kj}\approx 0$, the power spectrum of $\zeta$ is incompatible between the two approaches, as shown in the upper panel in Fig ~\ref{fig:SpecModel2}. This discrepancy is mainly due to the $k^2$ correction which is neglected in the separate universe approach but is now as large as the leading order contribution. As we choose later initial hypersurfaces, i.e., larger $N_{kj}$'s, the power spectra given by the ordinary $\delta N$ formalism approach the correct result obtained by numerically solving the Mukhanov-Sasaki equation. On the other hand, in our extended $\delta N$ formalism, we take into account the spatial curvature on the initial hypersurface, which significantly alleviates the discrepancy even if we set the initial condition as early as the horizon-exit moment.
This is clearly shown in the lower panel in Fig.~\ref{fig:SpecModel2}. 
We can also see some small discrepancies from higher orders of $k^2$ for $N_{kj}=0$, which disappears rapidly and monotonically as we increase $N_{kj}$.

\textit{Non-Gaussianities}---A nonlinear treatment of these models is demanded, as non-Gaussianities of density perturbations are crucial to predict the production rate of primordial black holes \cite{Franciolini:2018vbk,Ezquiaga:2018gbw,Firouzjahi:2018vet,Yoo:2019pma,Kehagias:2019eil,Ballesteros:2020sre,Cai:2021zsp,Riccardi:2021rlf,Taoso:2021uvl,Biagetti:2021eep,Kitajima:2021fpq,Pattison:2021oen,Cai:2022erk,Young:2022phe,Escriva:2022pnz,Escriva:2022duf,Matsubara:2022nbr,Ferrante:2022mui,Gow:2022jfb,LISACosmologyWorkingGroup:2023njw,Tomberg:2023kli,Ianniccari:2024bkh,Pi:2024jwt}.
Here, we briefly discuss how to evaluate the nonlinear parameter $f_{\rm NL}$ by  applying the extended $\delta N$ formalism, but defer more detailed discussion about the non-Gaussian probability distribution of the perturbation to future work.

By denoting $\zeta_G$ the Gaussian curvature perturbation, deviations from this Gaussianity can be captured by the quadratic term with a nonlinear parameter $f_{\rm NL}$ \cite{Komatsu:2001rj,Maldacena:2002vr,Bartolo:2004if,Yokoyama:2008by},
\begin{align}
    \zeta = \zeta_G + \frac{3}{5} f_{\rm NL} \zeta_G^2\,+\cdots. \nonumber
\end{align}
Defining $\zeta_1=\partial\zeta(N)/\partial\zeta(N_j)$, $\zeta_2=\partial\zeta(N)/\partial\zeta_N(N_j)$, $\zeta_{12}=\partial^2\zeta(N)/\partial\zeta(N_j)\partial\zeta_N(N_j)$, etc., we can write $f_{\rm NL}$ in the following expression~\cite{Yokoyama:2007dw}
\begin{align}
   f_{\rm NL} = \frac{5}{6}   \frac{\zeta_{a} \zeta_{b} \zeta_{cd}P^{ac} P^{bd}} {\left(\zeta_{e} \zeta_{f}  P^{ef}\right)^2} 
   \,; \label{fnl}
\end{align}
we use Einstein summation convention for $a,\dots, f=1,2$, and we assume that the initial distribution of $\zeta(N_j)$ and $\zeta_N(N_j)$ is Gaussian. $P^{ab}$ are the two-point correlation functions of $\zeta\left(N_j\right)$ or $\zeta_N\left(N_j\right)$ evaluated by using the standard perturbation theory of a Bunch-Davies vacuum state.
Concerning the ultra-slow-roll stage, for simplicity, we set $U_\phi^{\rm II}=0$ for the following analytic computation. In this case, Eq.~\eqref{N12} simplifies to 
\begin{align}\nonumber
    N_{12} = - \frac{1}{3}\ln\beta ,~\text{with}~
 \beta:=1 - \frac{3\phi_{12}}{\phi_N(N_1)} - \frac{\mathcal{K}e^{-2N_{j1}}}{5 H_0^2} \,,
\end{align}
where $\phi_N(N_1)$ and $N_{j1}$ are obtained from Eqs.~\eqref{phi0}--\eqref{phi1}, their derivatives, and Eq.~\eqref{Nj1}, which depend on 
$\phi(N_j)$, $\phi_N(N_j)$ and ${\cal K}$. Roughly speaking, $\beta$ is approximately given by $\phi_{2v}/\phi_{1v}$, where $\phi_v$ would mark the endpoint of the ultra-slow-roll evolution if $U^{\rm II}$ were to persist beyond $\phi_2$.
Let's assume that the contribution of $\delta N_{12}$ dominates in $\zeta(N_2)$. 
Then, we expand $\beta$ between background and perturbations, respectively, as 
$\beta=\overline{\beta}+\delta \beta+ O(\zeta(N_j)^2, \zeta(N_j) \zeta_N(N_j),\zeta_N(N_j)^2)$, 
where $\bar\beta$ is the background value of $\beta$. In ultra-slow-roll inflation, the enhancement of $\delta\beta/\bar{\beta}$ is realized by the smallness of $\bar\beta$, hence we can neglect the higher-order terms and $\delta\beta$ is a linear 
function of $\zeta(N_j)$ and $\zeta_N(N_j)$, whose distributions can be well approximated by Gaussian distributions. 
In this case, Eq.~\eqref{fnl} can be reduced to  
\begin{align}
        f_{\rm NL} = \frac{5}{6}  
    \frac{\partial_{\bar\beta}^2N_{12}({\bar\beta}) }
    {\left(\partial_{\bar\beta} N_{12}({\bar\beta})\right)^2} \,.
\end{align}
Then, it is easy to show that the nonlinear parameter reduces to $f_{\rm NL}=5/2$. Zero or negative $\bar\beta$ corresponds to infinite $e$-folding number as the inflaton gets stuck on the plateau, and quantum diffusion is needed to end inflation \cite{Ezquiaga:2018gbw,Firouzjahi:2018vet,Ballesteros:2020sre,Pattison:2021oen,Tomberg:2023kli}, which is beyond our scope.

\textit{Conclusion}---The $\delta N$ formalism is a nonlinear approach allowing one to compute the curvature perturbation $\zeta$ by the perturbed $e$-folding number $\delta N$ in a perturbed FLRW universe.
This method relies on the separate-universe approach which captures the superhorizon-scale dynamics by neglecting gradient terms of $\mathcal{O}\left(k^2\right)$. Effectively, it is equivalent to evolving independently a set of causally disconnected patches, 
each of which is a flat, homogeneous and isotropic FLRW universe. In certain scenarios, however, the adiabatic mode may exhibit important gradient corrections of ${\cal O}(k^2)$, which leads to the breakdown of the separate-universe picture. One important example is the model with an ultra-slow-roll phase, which is among the scenarios that introduce a peak in the power spectrum. In this Letter, we showed how to capture the $k^2$ corrections of the adiabatic mode within the framework of the $\delta N$ formalism by introducing the spatial curvature $\mathcal{K}$ in each patch of the separate universe. The initial conditions of the separate universe are identified by matching with linear-perturbation theory, and in this extended $\delta N$ formalism the curvature $\mathcal{K}$ takes care of the $k^2$ correction in $\zeta$. 
Namely, moving to the gauge in which the inflaton field takes a constant value on the equal-time hypersurface at the initial time, we absorb the spatial 
gradient terms into the spatial curvature of the hypersurface. By doing so, the gradient term in the Klein-Gordon equation is made irrelevant for the adiabatic mode, and one can accurately compute $\zeta$ even if the separate-universe approach is used right after 
the horizon exit. We illustrated this methodology in the case of a Starobinsky model and confirmed the validity of this method by explicitly comparing the resulting power spectrum of $\zeta$ with the numerical result of the linear perturbation theory. A formal proof of the validity of this method is presented in the \textit{end matter}, and an analytic comparison with the linear-perturbation theory is put in Supplemental Material.
Finally, we used the extended $\delta N$ formalism to compute the non-Gaussianities of the curvature perturbation. We observed that the $f_{\rm NL}$ parameter value can make a plateau at $f_{\rm NL}=5/2$ and the plateau would contain the peak frequency of the power spectrum if the ultra-slow-roll phase abruptly transits to another slow-roll phase.
From the analytic estimate, on the plateau the distribution of $\zeta$ is determined from a Gaussian distribution of $\delta {\beta}$ by the nonlinear transform that takes the form of $\zeta=-\frac13 \ln \left(1+\delta\beta/\bar\beta\right)$. 
Hence, the distribution at a large positive value of $\zeta$ behaves like $\propto \exp(-3\zeta)$~\cite{Pi:2024jwt}. 
To give the full frequency dependence of $f_\mathrm{NL}$, we need more careful treatment, which we would like to defer to future work.

{\it Acknowledgments}---
We thank Diego Cruces, Adrian Palomares, Misao Sasaki, and David Wands for discussions and useful comments. 
This work is supported in part by the National Key Research and Development Program of China Grant No. 2021YFC2203004. The authors thank the YITP long-term workshop ``Gravity and Cosmology 2024," during which the basic idea of this work has been developed. 
D.~A. is supported by JSPS Grant-in-Aid for Scientific Research Grant No. JP23KF0247.
S.~P. is supported by National Natural Science Foundation of China Grants No. 12475066 and No. 12447101, by JSPS KAKENHI Grant No. JP24K00624, and by the World Premier International Research Center Initiative (WPI Initiative), MEXT, Japan.
T.~T. is supported by Grant-in-Aid for Scientific Research under Contract No. JP23H00110, No. JP20K03928, No. JP24H00963, and No. JP24H01809. 
\bibliographystyle{unsrturl}
\bibliography{deltaN}

\newpage
\onecolumngrid
\appendix
\newpage
\section{End matter:\\ Gradient expansion with local curvature} \label{app:MatchingConditions}

\twocolumngrid
\subsection{Initial conditions from linear perturbations}

Here we focus on scalar-type perturbations. The metric of the scalar-type perturbation can be written as \cite{Bardeen:1980kt,Kodama:1984ziu,Sasaki:1998ug}, 
\begin{align}\nonumber
ds^2&=a^2\Big[-\big(1+2AY\big)d\eta^2-2BY_jd\eta dx^j\\\label{metric0}
&\quad\quad+\big((1+2H_L)Y\delta_{ij}+2H_TY_{ij}\big)dx^idx^j\Big]\,.
\end{align}
where $Y$ is the spatial scalar harmonic with the eigenvalue $k^2$, and
\begin{align}\nonumber
Y_j &= -\frac1k\nabla_j Y, \\\nonumber
Y_{ij} &= \frac1{k^2}\left(\nabla_i\nabla_j Y+\frac13\delta_{ij}\nabla^2Y\right)\,. 
\end{align}
We associate the harmonics $Y$ explicitly to emphasize that the metric and scalar-field perturbations are to be understood as the expansion coefficients here, although we use the same notations to express the corresponding spacetime functions.
The local expansion along a geodesic is
\begin{equation}
\tilde{N}=\int^{\eta}_{\eta_0}\left(\mathcal{H}+\Big(H_L'+\frac13kB\Big)Y\right)d\eta\,,
\end{equation}
which implies the $e$-folding number $\tilde{N}$ equals to the background $N$ if we take the $\delta N$ gauge
\begin{align}
H_L'=B=0\,.
\end{align}
This gives a constraint for the curvature perturbation, $\mathcal{R}\equiv H_L+\frac13 H_T$, 
\begin{align}\label{R'constraint}
\mathcal{R}'=\frac13H_T'\,.
\end{align}
It is easy to see that there is some gauge redundancy hidden in the integration constant of \eqref{R'constraint}, which we will use later to set the initial conditions of $\delta\phi$.

The $\delta N$ gauge is convenient to see the equivalence of the perturbed equation and the background equation \cite{Sasaki:1998ug}. In this gauge, the perturbed Klein-Gordon equation at linear order reduces to 
\begin{align}\nonumber
  H{d\over dN}\Big(H {\delta\phi}_N\Big)
 +3H^2{\delta\phi}_N +\Upp {\delta\phi} &\\
 +2\Up A - H^2{\phi_N} A_N + k^2 e^{-2N} {\delta\phi} &=0 \,. 
\label{chieq}
\end{align}
Thus among metric variables, the perturbed field equation contains only $A$. From the $\left({}^0_0\right)$ component of the perturbed Einstein equations, one can see that $A$ is expressed in terms of $\delta\phi$ as 
\begin{align}
 2 U A=-H^2 \phi_N {\delta\phi}_N-\Up{\delta\phi}+2k^2 e^{-2N} \calR\,. 
\label{Aeq}
\end{align}
At this point, if one can neglect the last term proportional to $k^2$, one may substitute Eq.~\eqref{Aeq} into 
Eq.~(\ref{chieq})
to obtain a closed second-order equation for ${\delta\phi}$. 
From the traceless part of the $\left({}^i_j\right)$ component of Einstein equations, the equation for ${\cal R}$ can be written under the closed form 
\begin{align}
 \frac1{a^3 H} \partial_N \left(a^3 H \partial_N{\cal R}\right)
   =\frac{k^2}{3a^2 H^2}({\cal R}+A) \,.  \label{DiffEqnR}
\end{align}
Now, we recall that the $k^2$ correction is necessary only for the adiabatic mode. At any reference time 
we can set $\delta\phi=0$, attributing all the curvature perturbation $\zeta$ to ${\cal R}$,  
which is possible because the $\delta N$ gauge is not a complete gauge fixing (see Eq.~\eqref{R'constraint}).
Moreover, from the nonadiabatic mode, one can also set ${\delta\phi}_N=O(k^2)$ at the reference time. 
Then, the above perturbation equations indicate that both $\delta\phi$ and $A$ remain $O(k^2)$. 
As a result, the last term in Eq.~\eqref{chieq} and the contribution of $A$ in Eq.~\eqref{DiffEqnR} become $O(k^4)$, hence providing us with a closed equation for $\mathcal{R}$.

\subsection{Separate-universe mapping}
In this subsection we derive the aforementioned equations for the separate universes, which are causally disconnected patches and evolve independently after $N_j$. Using the number of $e$-folds $N$ as the time coordinate, the Klein-Gordon equation and 
the FLRW equation with spatial curvature $\mathcal{K}$ defined at some initial time $N_j$ become 
\begin{eqnarray}
&& H{d\over dN}\left(H{\phi_N}\right)
  +3H^2{\phi_N}+\Up=0\,, 
\label{bgNK}
\\
&& H^2 \left(1-{1\over 6}\phi_N^2\right)={1\over 3}U-\mathcal{K} e^{-2\left(N-N_j\right)}\,. 
\label{frNK}
\end{eqnarray}
We will now assume $\mathcal{K}=0$ in the unperturbed background (or equivalently, in the fiducial universe) such that $\mathcal{K}$ is thought to be first order in perturbations. For a perturbed universe, the variation of Eqs.~\eqref{bgNK} and \eqref{frNK} yields
\begin{widetext}
\begin{eqnarray}
&& H{d\over dN} \left(H \delta\phi_{N}\right)
  +3H^2\delta \phi_{N}+\Upp\delta \phi
-2\Up{\delta H\over H} 
+ H^2{\phi_N}{d\over dN}\left({\delta H\over H}\right)=0\,,
\label{philameq}\\
 &&-2U{\delta H\over H}
   =-{H^2}\phi_N \delta \phi_{N}
     - \Up \delta \phi + 3e^{-2\left(N-N_j\right)} \mathcal{K} \,,
\label{Hlameq}
\end{eqnarray}
\end{widetext}
We find these equations are, respectively, equivalent to  Eqs.~(\ref{chieq}) and (\ref{Aeq})
in the $\delta N$ gauge,
with the identifications
\begin{align}
    \frac{\delta H}{H}&=-A\,,\\\label{K-krelation} 
    e^{2 N_j} \mathcal{K}&=\frac{2}{3}k^2{\cal R}\,, 
\end{align}
except for the last term on the left-hand side of Eq.~\eqref{chieq}. Notice that for a piecewise-linear potential, $\Upp$ vanishes in each segment and the linear equations \eqref{chieq} and \eqref{Aeq} become exact, thus allowing for a nonlinear analysis. Equation~\eqref{K-krelation} can also be obtained in the following way. 
Note that on a homogeneous isotropic equal-time hypersurface with a curvature term $\mathcal{K}e^{-2(N-N_j)}$, the three-dimensional curvature is equal to $6\mathcal{K}e^{-2(N-N_j)}$, while from the metric Eq.~\eqref{metric0}, this curvature is $-4e^{-2N}\nabla^2\mathcal{R}$. Then, we easily check the consistency of Eq.~\eqref{K-krelation}.

As mentioned above below Eq.~\eqref{DiffEqnR}, we do not need the term proportional to $k^2$ in Eq.~\eqref{chieq} to obtain the adiabatic mode in an appropriate choice of the residual gauge degrees of freedom.
This proves that the separate-universe description can completely reproduce the linear perturbation including the $k^2$ correction of the adiabatic mode, if we set the matching conditions appropriately. 
An important point is that the gauge-invariant comoving curvature perturbation $\zeta$ should be attributed to ${\cal R}$ as an initial condition for the separate-universe evolution. Otherwise, the last term $k^2e^{-2N}\delta\phi$ in Eq.~\eqref{chieq}, which is missing in Eq.~\eqref{philameq}, contributes as the correction of $O(k^2)$.

The initial condition at $N=N_j$ should be provided in terms of $\zeta$ as follows:
\begin{align}
 \zeta(N_j)&=\left.{\cal R}(N_j)-\frac{\delta \phi}{\phi_N}\right|_{Nj}\,,\\\label{app:zetaN}
 \zeta_N(N_j)&=\mathcal{R}_N(N_j)-\partial_N\left(\frac{\delta \phi}{\phi_N}\right)_{N_j}\,,\\
  \delta\phi(N_j)&=0\,.
\end{align}
The three equations above cannot determine four variables, ${\cal R}$, $\delta\phi$, and their derivatives.  
We need to supplement the condition coming from the momentum constraint, i.e., the $\left({}^0_i\right)$ component of the perturbed Einstein equations
\begin{equation}
 \calR_N=A-{1\over 2}\phi_N {\delta\phi}\,.
\label{0ieq}
\end{equation}
Combined with Eq.~\eqref{Aeq}, we eliminate $A$ to obtain another relation among the variables to be determined, 
\begin{equation}\label{app:intermediate1}
2U\mathcal{R}_N(N_j)=-H^2\phi_N(N_j){\delta\phi}_N(N_j)+2k^2e^{-2N_j}\mathcal{R}(N_j)\,,
\end{equation}
where we have used $\delta\phi(N_j)=0$. Substituting \eqref{app:zetaN} into \eqref{app:intermediate1}, we can eliminate $\mathcal{R}_N(N_j)$ and obtain
\begin{equation}\label{app:initConds2}
\delta\phi_N(N_j)=\frac{\phi_N}{3H_0^2}\left(-U\zeta_N(N_j)+k^2e^{-2N_j}\zeta(N_j)\right)\,.
\end{equation}
Then substituting \eqref{app:initConds2} back into \eqref{app:zetaN}, we can easily derive the condition for $\mathcal{R}_N(N_j)$, shown in Eqs.~\eqref{initConds}.

Neglecting the contribution of $A$ in Eq.~\eqref{DiffEqnR}, 
one can solve the equation to determine the leading $k^2$ correction contained in $\mathcal{R}$. If we allow to approximate $H_0$ to be constant, we get
\begin{align}\label{R-k^2}
    \mathcal{R}(N) &= 
    \mathcal{R}\left(N_j\right) \left[1+\frac{k^2}{6 H_0^2} 
e^{-2 N_j} 
    \left(1-e^{-2\left(N-N_j\right)}\right) 
    \right]\,,
\end{align}
where we used the initial condition for $\mathcal{R}_N$ given in Eqs.~\eqref{initConds}, neglecting the contribution from $\zeta_N$ at $N=N_j$.
After a few $e$-folds,
\begin{align}
    \mathcal{R} (N) &\approx \left[1+\frac{k^2}{6 H_0^2} e^{-2 N_j}\right] \mathcal{R}\left(N_j\right) \,. 
\end{align}
This solution clearly indicates that the $k^2$ correction in ${\cal R}$ remains approximately constant and does not have any enhancement factor due to the ultra-slow roll phase.

\newpage
\onecolumngrid

\section{Supplemental Material: \\ Linear approximation in ultra-slow-roll inflation} \label{app:Perturb-gdN}
\subsection{Extended $\delta N$ for the Starobinsky model}
\subsubsection{Modes crossing during slow roll}
For an application, we detail the calculations of the extended $\delta N$-approach in the context of linear (field) perturbations. As mentioned earlier we will assume that the background curvature vanishes such that $\mathcal{K}$ is of first-order in perturbative expansion. This means that the perturbations of the scalar field are given by
\begin{align}
    \delta\phi &:= \delta \phi^{(0)} + \phi^{(1)}\,,
\end{align}
where we recall that in our notations, the upper index refers to the order in $\mathcal{K}$ expansion.

We start our analysis with the case where the extended $\delta N$ is matched to linear-perturbation theory during the slow-roll phase, $N_j < N_1$. From Eqs.~\eqref{phi0} and \eqref{phi1}, the condition $\phi_1\equiv\phi(N_1)$ is explicitly written down as
\begin{align}
 \phi_1 &\approx \phi(N_j) -\frac{U^{\rm I}_\phi}{3H_0^2} \DeltaN_{j1} -\frac13 \left(\phi_{N}(N_j)+\frac{U^{\rm I}_\phi}{3H_0^2}\right)
      \left(e^{-3\DeltaN_{j1}}-1\right)  -  \frac{ {\cal K}U^{\rm I}_\phi}{9 H_0^4} \left(1-3e^{-2\DeltaN_{j1}}\right)  \,,
\label{eqforN1}
\end{align}
where we kept the leading order in $\mathcal{K}$, neglected the remaining terms that decay exponentially fast, and used the slow-roll condition at $N_j$ to rewrite $\mathcal{K} \phi^{(0)}_N(N_j) = - \mathcal{K} \Up^{\rm I}/(3H_0^2)$.

We fix the value $\phi_1$ at the transition such that, at linear order, $\delta\phi_1=0$. Expanding the above equation in perturbations, the perturbed $e$-folding number between the start and the end of the first slow-roll phase is approximated by
\begin{equation}
 \delta \DeltaN_{j1}\approx \frac{3H_0^2}{U^{\rm I}_\phi}\left(\delta \phi(N_j)
     +\frac13\delta\phi_{N}(N_j) \right)
    -\frac{{\cal K}}{3H_0^2} \left(1-3e^{-2\bar\DeltaN_{j1}}\right)\,,
\end{equation}
where we neglect terms decaying as $e^{-3 N_{j1}}$ and denote by $\bar N$ the background $e$-folding number (notice that $N_j$ is unperturbed by definition so we do not put overline to $N_j$ below).
We focus on the adiabatic mode, whose $k^2$ correction is relevant. Hence, setting $\zeta'(N_j)=0$, the junction conditions Eqs.~\eqref{initConds}, which include $\delta\phi(N_j)=0$, lead us to
\begin{equation}
 \delta N_{j1}\approx
    -\frac{{\cal K}}{2H_0^2} + \frac{\cal K}{H_0^2} e^{-2 \bar N_{j1}} \approx - \frac{k^2 \zeta_j}{3 H_0^2} e^{-2 N_j} + \frac{2 k^2 \zeta_j }{3 H_0^2} e^{-2 \bar N_1} \,. 
\end{equation}
In the first equality, we neglect the term decaying like $e^{-3N_1}$. 
The comoving curvature perturbation $\zeta$ at $N_1$ at linear order is therefore evaluated by
\begin{align}
    \zeta_{j1} &= \delta N_{j1} +\frac{ k^2 \zeta_j}{6 H_0^2} \left(e^{-2N_j}-e^{-2\bar N_1}\right) \approx - \frac{k^2 \zeta_j}{6 H_0^2} e^{-2 N_j} + \frac{ k^2 \zeta_j}{2 H_0^2} e^{-2\bar N_1} \,, 
 \label{deltaN1}
\end{align}
with the $k^2$ correction also provided by linear-perturbation theory, Eqs.~\eqref{R-k^2}. Upon neglecting the term decaying like $e^{-3N_1}$, this expression matches the one found from a linear-perturbation approach, see Eq.~\eqref{deltaNj1-LinearPert} below.

We then study the following ultra-slow-roll phase. The initial condition at the junction time $N_1$ can be specified by the continuity of the solution Eqs.~\eqref{phi0} \& \eqref{phi1} together with their $N$-derivative:
\begin{align}
    \phi^{(0)}_N &= - \frac{U_\phi}{3 H_0^2} + \left(\phi^{(0)}_N(N_*) + \frac{U_\phi}{3 H_0^2}\right) e^{-3\left(N-N_*\right)} \,, \label{phi0N} \\
    \phi^{(1)}_N &= \phi^{(1)}_N(N_*)e^{-3\left(N-N_*\right)}  + \frac{\mathcal{K} U_\phi}{6 H_0^4} e^{-2\left(N-N_j\right)} \left[ -4 + 3 e^{-\left(N-N_*\right)} + e^{-3\left(N-N_*\right)} \right]  \label{phi1N}\\
    & - \frac{\mathcal{K} \phi^{(0)}_N (N_*)}{2 H_0^2} e^{-2\left(N-N_j\right)} \left[  e^{-\left(N-N_*\right)} -  e^{-3\left(N-N_*\right)} \right] \,.  \nonumber
\end{align}
The field $\phi_1$ is unperturbed by construction, while the perturbations of its derivative can be approximated by
\begin{align}
\phi_N(N_1) &= - \frac{U^{\rm I}_\phi}{3 H_0^2} + \left(\phi_N(N_j) + \frac{U^{\rm I}_\phi}{3 H_0^2}\right) e^{-3\DeltaN_{j1}} - \frac{2 \mathcal{K} U^{\rm I}_\phi}{3 H_0^4} e^{-2 \DeltaN_{j1}} - \frac{\mathcal{K} \phi^{(0)}_N(N_j)}{2 H_0^2} e^{-3 \DeltaN_{j1}} \,.
\label{phiNN1}
\end{align}
The field $\phi$ can then be propagated during the ultra-slow-roll phase by rewriting Eqs.~\eqref{phi0} and \eqref{phi1} for $N_*=N_1$. The condition $\phi_2\equiv\phi(N_2)$ reads
\begin{align}
    \phi_2&= \phi_1 -\frac{U^{\rm II}_\phi}{3H_0^2} N_{12} -\frac13 
     \left( \phi_{N}(N_1) +\frac{U^{\rm II}_\phi}{3H_0^2}\right)
        \left(e^{-3N_{12}} -1\right) +  \frac{{\cal K}U^{\rm II}_\phi}{3H_0^4} e^{-2 N_{j1}} \left(- \frac{2}{5} + e^{-2 N_{12}} \right) - \frac{\mathcal{K} \phi^{(0)}_N(N_1)}{15 H_0^2} e^{-2 N_{j1}} \,,
\end{align}
where we neglect terms of order $\mathcal{K}$ decaying like $e^{-3N_{12}}$ or $e^{-5 N_{12}}$. By massaging the above expression, we find that, at first order in perturbations,
\begin{align}
    \left(\frac{e^{-3 \bar \DeltaN_{12}} + \hat U}{\hat U}\right)\delta N_{12}&\approx \frac{3 H_0^2}{U^{\rm II}_\phi} \left[- \delta \phi_{12} - \frac{1}{3} \left(e^{-3 \bar \DeltaN_{12}}-1\right) {\delta\phi}_N(N_1) + \frac{ \mathcal{K} U^{\rm II}_\phi}{3 H_0^4} e^{-2\bar\DeltaN_{j1}}\left(-\frac{2}{5}+e^{-2\bar N_{12}}\right)  + \frac{{\cal K} U^{\rm I}_\phi}{45 H_0^4} e^{-2\bar \DeltaN_{j1}}\right] \,, \nonumber \\
\end{align}
where we used the slow-roll condition $\mathcal{K} \phi^{(0)}_N(N_1) = - \mathcal{K} U^{\rm I}_\phi/ \left(3H_0^2\right)$ and defined $\hat U:=U^{\rm II}_\phi/(U^{\rm I}_\phi-U^{\rm II}_\phi)$. Perturbing Eq.~\eqref{phiNN1}, we find that
\begin{align}
    {\delta\phi}_N(N_1) &\approx - \frac{2 \mathcal{K} U^{\rm I}_\phi}{3 H_0^4} e^{-2 \bar \DeltaN_{j1}} \,,
\end{align}
where we neglected terms decaying as $e^{-3\DeltaN_{j1}}$. Plugging this in the equation for $\delta N_{12}$ and noticing that $\delta\phi_{12}=0$ by definition, we can use the initial condition of $\mathcal{K}$, Eqs.~\eqref{initConds}, to rewrite
\begin{align}
   \delta N_{12} &\approx \frac{\zeta_j e^{-2 \bar N_1}}{e^{-3 \bar \DeltaN_{12}} + \hat U} \left(\frac{1}{U^{\rm I}_\phi-U^{\rm II}_\phi}\right) \left[- \frac{2 k^2 U^{\rm I}_\phi}{5H_0^2} + \frac{2 k^2 U^{\rm II}_\phi}{ 3 H_0^2}\left(-\frac{2}{5}+e^{-2\bar N_{12}}\right) \right]\,.
\end{align}
Using the $k^2$ correction in $\mathcal{R}$, Eq.~\eqref{R-k^2}, the change of the comoving curvature perturbation from $N_1$ to $N_2$ boils down to
\begin{align}
    \zeta_{12} &= \delta N_{12} +\frac{ k^2 \zeta_j}{6 H_0^2} \left(e^{-2\bar N_1}-e^{-2\bar N_2}\right) = \frac{\zeta_j e^{-2 \bar N_1}}{e^{-3 \bar \DeltaN_{12}} + \hat U} \left(\frac{1}{U^{\rm I}_\phi-U^{\rm II}_\phi}\right) \left[- \frac{2 k^2 U^{\rm I}_\phi}{5H_0^2} - \frac{ k^2 U^{\rm II}_\phi}{10 H_0^2} + \frac{k^2 U^{\rm II}_\phi}{2 H_0^2} e^{-2\bar N_{12}} \right]\,. 
    \label{deltaN12-Starobinsky}
\end{align}
This indeed matches the equation found below in linear perturbations~\eqref{deltaN12-LinearPert}.

As an additional check, one can take the limit $U^{\rm I}_\phi=U^{\rm II}_\phi$ and notice from Eq.~\eqref{deltaN1} that
\begin{align}
    \zeta_{j2} = - \frac{k^2 \zeta_j }{6 H_0^2}e^{-2  N_j} + \frac{k^2 \zeta_j}{2 H_0^2} e^{-2 \bar N_2} \,.
\end{align}
This coincides with the result expected for a continuous slow-roll phase from $N_j$ to $N_2$, which is equivalent to replacing $\bar N_1$ with $\bar N_2$ in Eq.~\eqref{deltaN1}.

\subsubsection{Modes crossing during ultra-slow roll}
For the modes that cross the horizon during the ultra-slow-roll phase, we evolve the scalar field from the matching time $N_j(>N_1)$. Upon setting $N_*=N_j$ in Eqs.~\eqref{phi0} and \eqref{phi1}, the condition $\phi_2\equiv\phi(N_2)$ becomes 
\begin{align}
    \phi_2 &= \phi_j -\frac{U^{\rm II}_\phi}{3H_0^2} N_{j2}
    -\frac13  \left( \phi_{N}(N_j) +\frac{U^{\rm II}_\phi}{3H_0^2}\right)
        \left(e^{-3N_{j2}}-1\right)   +  \frac{{\cal K}U^{\rm II}_\phi}{3H_0^4} \left(-\frac{2}{5} + e^{-2 N_{j2}}\right) - \frac{\mathcal{K}\phi^{(0)}_N\left(N_j\right)}{15 H_0^2} \,,
\end{align}
where we neglected $\mathcal{K}$-terms decaying as $e^{-3 N_{12}}$ and $e^{-5 N_{12}}$. 
Using Eq.~\eqref{phiN-LinearPert}, the initial condition for the background value of $\phi_N\left(N_j\right)$, 
the perturbed number of $e$-folds at linear order is therefore
\begin{align}
        \left(\frac{e^{-3 \bar \DeltaN_{12}} + \hat U}{\hat U}\right) \delta N_{j2}&\approx \frac{3 H_0^2}{U^{\rm II}_\phi} \left[- \delta \phi_{j2} - \frac{1}{3} \left(e^{-3 \bar \DeltaN_{j2}}-1\right) {\delta\phi}_N(N_j) - \frac{ 2\mathcal{K} U^{\rm II}_\phi}{15 H_0^4}  - \frac{{\cal K} \phi_N^{(0)}\left(N_j\right)}{15 H_0^2} \right] \,, 
\end{align}
when considering only the leading-order terms at order $\mathcal{K}$. 
Recalling $\delta\phi_{j2}=0$ and plugging the initial condition for $\delta \phi_N$ in Eqs.~\eqref{initConds}, this equation becomes
\begin{align}
    \delta N_{j2} &= - \frac{k^2 \zeta_j}{15 H_0^2} \frac{ e^{3 \bar N_{j1}} + 5 \hat{U}}{e^{-3 \bar N_{12}}+\hat{U}} e^{-2 N_j} \,.
\end{align}
We now add the $k^2$ correction in $\mathcal{R}$, Eq.~\eqref{R-k^2},
to obtain
\begin{align}
    \zeta_{j2} = \delta N_{j2} + \frac{k^2 \zeta_j}{6 H_0^2} \left(e^{-2 N_j} - e^{-2 \bar N_2}\right) \approx - \frac{k^2 \zeta_j}{H_0^2} \left( \frac{e^{3 \bar N_{j1}}}{15} + \frac{\hat{U}}{6} \right)  \frac{e^{-2 N_j}}{e^{-3 \bar N_{12}} + \hat{U}} \,,
\end{align}
where in the brackets of the right-hand side we neglected terms decaying as $e^{- N_2}$ in the numerator. This result matches the calculation from perturbation theory, given in Eq.~\eqref{deltaN12-USRmodes} below.

\subsection{$k^2$ correction in linear perturbation} \label{sec:LinPert}

\subsubsection{Modes crossing during slow roll}

We estimate the $k^2$ correction to the adiabatic mode of $\zeta$ in the context of linear-perturbation theory. We start our analysis in conformal time $\eta$ which is more standard. Since in general $e^{-N}/H_0=\eta$, the shape of the Starobinsky potential gives us the background field velocity, which we denote with an overbar as follows:
\begin{equation}
    \bar \phi_N=\left\{
     \begin{array}{ll}
     \displaystyle -\frac{U^{\rm I}_\phi}{3H_0^2}\,,\qquad &(\eta\leq\eta_1,~~\mbox{segment I})\,,\cr
     \displaystyle -\frac{U^{\rm I}_\phi-U^{\rm II}_\phi}{3H_0^2}\left(\frac{\eta}{\eta_1}\right)^{\!\!3}-
      \frac{U^{\rm II}_\phi}{3H_0^2}\,,\qquad &(\eta\geq\eta_1,~~\mbox{segment II})\,. \label{phiN-LinearPert}
      \end{array}\right.
\end{equation}
We remind that $z\left(\eta\right)=a\left(\eta\right)\phi_N\left(\eta\right)=a(\eta_1) \,\phi_N\left(\eta\right)(\eta_1/\eta)$, where $\eta_1:=e^{-N_1}/H_0$. Since $N$ always appears as a background value in this section, we do not associate overbar, for simplicity. We start looking at the case where the $\delta N$ formalism is matched to linear-perturbation theory at some time $\eta_j(<\eta_1)$. Upon neglecting the nonadiabatic mode, the curvature perturbation at $\eta_2:=e^{-N_2}/H_0$ is given by
\begin{align}
 \zeta\left(\eta_2\right) \approx \zeta_j \, u_\text{ad}(\eta_2) &\approx  
 \zeta_j- k^2\zeta_j \int_{\eta_j}^{\eta_2} 
   \frac{d\eta }{z^2(\eta)}\int_{\eta_j}^{\eta} d\eta' \, z^2(\eta')\,.
\end{align}
The integral can then be split between the slow-roll and ultra-slow-roll phases
\begin{align}
\zeta_{j2} &\approx  \underbrace{- k^2\zeta_j \int_{\eta_j}^{\eta_1} d\eta \,\eta^2
     \int_{\eta_j}^{\eta} \frac{d\eta'}{\eta'{}^2}}_{\zeta_{j1}}
     \underbrace{- k^2\zeta_j  \int_{\eta_1}^{\eta_2} 
   \frac{d\eta \,\eta^2}
      {\displaystyle\left[\left(\frac{\eta}{\eta_1}\right)^{\!\!3}+\hat U \right]^2}
     \left(\int_{\eta_j}^{\eta_1} \frac{d\eta'}{\eta'{}^2} \left[1 + \hat{U}\right]^2 +
     \int_{\eta_1}^{\eta} \frac{d\eta'}{\eta'{}^2} \displaystyle\left[\left(\frac{\eta'}{\eta_1}\right)^{\!\!3}+\hat U \right]^2\right)}_{\zeta_{12}} \,.\nonumber \\
\end{align}
The part $\zeta_{j1}$ describes the first slow-roll evolution from $\eta_j$ to $\eta_1$, while the part denoted by $\zeta_{12}$ captures the ultra-slow-roll evolution from $\eta_1$ to $\eta_2$. The first slow-roll part is easily computed as
\begin{align}
    \zeta_{j1}  &= - \frac{k^2 \zeta_j}{H_0^2} e^{-2N_j} \left[\frac{1}{6} - \frac{1}{2} e^{-2N_{j1}} + \frac{1}{3} e^{-3N_{j1}} \right] \,.
     \label{deltaNj1-LinearPert}
\end{align}
Upon neglecting the decaying terms proportional to $e^{-2N_1}$ or $e^{-3N_1}$, this reduces to the expression we had found from the extended separate-universe approach~\eqref{deltaN1}.

On the other hand, the computation of the ultra-slow-roll part yields
\begin{align}
     \zeta_{12} &\approx - \frac{k^2 \zeta_j}{H_0^2} \left(\frac{2}{5}+\frac{\hat{U}}{2}\right)
       \frac{\eta_1^2}
      {\displaystyle\left(\frac{\eta_2}
     {\eta_1}\right)^{\!\!3}+\hat U}  + \frac{k^2 \zeta_j\hat{U}}{2 H_0^2}
       \frac{\eta_2^2}
      {\displaystyle\left(\frac{\eta_2}
     {\eta_1}\right)^{\!\!3}+\hat U} \nonumber \\
     &\approx -
       \frac{k^2 \zeta_j}{H_0^2} \frac{\eta_1^2}
      {\displaystyle\left(\frac{\eta_2}
     {\eta_1}\right)^{\!\!3}+\hat U} \left(\frac{1}{U^{\rm I}_\phi-U^{\rm II}_\phi}\right)\left(\frac{2}{5}U^{\rm I}_\phi +\frac{1}{10}U^{\rm II}_\phi\right) + \frac{k^2 \zeta_j\hat{U}}{2 H_0^2}
       \frac{\eta_2^2}
      {\displaystyle\left(\frac{\eta_2}
     {\eta_1}\right)^{\!\!3}+\hat U} \,,
     \label{deltaN12-LinearPert}
\end{align}
where we approximated the result using $|\eta_j| \gg |\eta_1| \gg |\eta_2|$ and defined $\hat U:=U^{\rm II}_\phi/(U^{\rm I}_\phi-U^{\rm II}_\phi)$. Under the case where $|U^{\rm I}_\phi| \gg |U^{\rm II}_\phi|$, going back to $e$-fold time, the result reduces to
\begin{align}
     \zeta_{12} &\approx -\frac{2 k^2 \zeta_j}{5 H_0^2}\frac{e^{-2N_1}}
      {\displaystyle e^{-3\DeltaN_{12}}+\hat U} \,, \label{deltaN12-LinearPert-U2=0}
\end{align}
and roughly agrees with the estimate by the extended $\delta N$ formalism~\eqref{deltaN12-Starobinsky}. If instead one poses $U^{\rm I}_\phi=U^{\rm II}_\phi$, then
\begin{align}
     \zeta_{12} &\approx - \frac{k^2 \zeta_j}{H_0^2} e^{-2N_1} \left[ \frac{1}{2} - \frac{1}{2} e^{-2N_{12}} \right] \,.
\end{align}
Adding this to the slow-roll counterpart~\eqref{deltaNj1-LinearPert}, we get
\begin{align}
     \zeta_{j2} = - \frac{k^2 \zeta_j}{H_0^2} e^{-2N_j} \left[\frac{1}{6}-\frac{1}{2}e^{-2N_{j2}}\right]\,,
\end{align}
and one finds the expected result from a continuous slow-roll expansion spanning from $N_j$ to $N_2$ up to a term decaying as $e^{-3N_{j2}}$ that we neglected. This can be quickly verified by replacing $N_1$ with $N_2$ in Eq.~\eqref{deltaNj1-LinearPert}.

\subsubsection{Modes crossing during ultra-slow roll}
Let us now analyse the case where the extended $\delta N$ formalism is used from some time $N_j(>N_1)$. In this context, the calculation from linear-perturbation theory gives us,
\begin{align}
     \zeta_{j2} &\approx - \frac{k^2 \zeta_j}{H_0^2} \left(\frac{e^{3 N_{j1}}}{15}+\frac{\hat{U}}{6}\right) \frac{e^{-2 N_j}}{e^{-3 N_{12}} + \hat{U}} \,, \label{deltaN12-USRmodes}
\end{align}
where we neglected additional terms decaying as $e^{- N_2}$ in the numerator.
\end{document}